\documentclass[12pt]{article}

\input amssym.def
\input epsf.sty

\begin{document}
\newtheorem{lemma}{Lemma}
\newtheorem{theorem}{Theorem}
\newtheorem{definition}{Definition}
\newcounter{one}
\setcounter{one}{1}
\newcounter{six}
\setcounter{six}{6}
\newcounter{nineteen}
\setcounter{nineteen}{19}
\renewcommand{\Re}{\mathop{\mathrm{Re}}}
\renewcommand{\Im}{\mathop{\mathrm{Im}}}

\begin{flushright}
\parbox{6cm}{
\begin{center}
This paper is dedicated \\
to the memory of Juergen Moser.
\end{center}
}
\end{flushright}

\begin{center}

{\large P.G.Grinevich}\footnote{Landau Institute for Theoretical
Physics of the Russian Academy of Sciences, Moscow 117940 Kosygin
Street 2, e-mail pgg@landau.ac.ru. This work is partially
supported by the RFBR Grant No~98-01-01161 and INTAS Grant
No~99-01782. Numerical simulations were partially performed on a
computer, donated to the author by the Humboldt Foundation.}
{\large S.P.Novikov}\footnote{IPST, University of Maryland-College
Park, MD 20742-2431 and Landau Institute for Theoretical Physics,
Moscow, e-mail novikov@ipst.umd.edu. This work is partially
supported by the NSF Grant DMS-0072700.}

\vspace{0.2cm}

{\bf\Large Topological Charge of the real periodic finite-gap Sine-Gordon
solutions}

\end{center}

\tableofcontents

\section{Introduction. Reality Problems for the Algebro-Geometric Solutions.}

The so-called finite-gap or algebro-geometric solutions were found for KdV in the
work \cite{NOV} in 1974. Their complete theory involving the spectral theory of
periodic (quasiperiodic) finite-gap Schr\"odinger operators on the line and its
unification
with algebraic geometry and analysis on the Riemann surfaces, time dynamics and
algebro-geometric hamiltonian aspects was constructed in the works
\cite{DN1,DN2,D1,IM1,L,McK-VM} (see in the survey article \cite{DMN}) and in the
works \cite{FMc,NV1}.

It was extended in many works to a huge number of 1+1 systems. In
the work \cite{K1} a purely algebraic procedure cleaned from the
spectral theory was formulated on the basis of these works and
applied to the $2+1$  systems like KP. The algebro-geometric
spectral theory of the stationary periodic 2D Schr\"odinger
operator was developed in the works \cite{DKN,NV}. Some classical
higher rank problems for the commuting OD operators were solved
(see \cite{KricNov}).

However, the algebraic methods free from the spectral theory have
a weak point in applications:  it is very difficult in many cases
to select effectively the proper ``real'' (i.e. physically or
geometrically meaningful) solutions among the complex solutions
given by the $\Theta$-functional expressions. Some physical
properties may be completely invisible from these  formulas. One
may observe that the effective solution of this problem is
normally easy exactly if corresponding Lax operator $L$ (or its
$\lambda$-dependent analog) is self-adjoint or has a good spectral
theory in the Hilbert space anyway, and difficult otherwise.

The reality problem is trivial for KdV. It is also trivial for the
defocusing Nonlinear Schr\"odinger equation $NLS_+$ and for Sinh-Gordon equation
$$
\frac{\partial^2 u}{\partial \xi \partial\eta}=4\sinh u(\xi,\eta).
$$

However, the reality problem is nontrivial for such famous classical systems
as the focusing Nonlinear Schr\"odinger $NLS_-$ and the Sine-Gordon (SG) system.
$$
u_{tt}-u_{xx}+\sin u(x,t)=0
$$

For example, until now there was no formula calculating such
elementary physical characteristic of the real finite-gap
(algebro-geometric) solutions periodic in the space variable $x$
with period $T$ as the so-called {\bf Topological Charge}. Let us
remind that we call function $u(x)$ periodic in this case if
$\exp\{iu(x+T)\} =\exp\{iu(x)\}$. The topological charge of this
function is equal to the integer number $n=[u(x+T)-u(x)]/2\pi$.
The density of the topological charge is equal to the real number
$n/T$. It can be extended to all quasiperiodic functions in the
variable $x$. This quantity survives all real periodic
nonintegrable perturbations. Therefore it is especially valuable
in applications.

Nobody was able to find out what the description of
the real components described in terms of the Jacoby torus and $\Theta$-functions
(found in the works \cite{Dubr-Nat,Erc-Forest}) implies for the values of the
topological charge.

{\bf\Large Problem}: How to calculate the topological charge of real periodic
SG solutions in terms of the Spectral Data: the Riemann surfaces and divisors.

An attempt has been made almost 20 years ago in the work \cite{DN} to develop
a method for the solution of this problem. This method has been called
an ``Algebro-Topological'' approach to the reality problems. An interesting formula
has been proposed. However, as it was pointed out
in \cite{Nov}, the idea of the proof presented in \cite{DN} failed. In this work we
found a new development of the main idea of the work \cite{DN}. We present here a
full proof of the formula for the topological charge based on the new effective
description of the real components in terms of divisors. A partial case of this
result is announced by the authors in the short note \cite{GN} where some
ideas are briefly described.

In the next work we are going to publish also a correction and a full proof
of the formula for the so-called Action Variables announced in the work \cite{DN}
(the standard field-theoretical Poisson Bracket restricted on the space
of finite-gap real SG solutions is the most interesting nontrivial case here).

\section{Finite-gap (algebro-geometrical) solutions to the Sine-Gordon
 equation. Known results.}

\subsection{The Sine-Gordon equation. Zero Curvature representation.}

The famous Sine-Gordon equation  (SG)

\begin{equation}
u_{tt}-u_{xx}+\sin u(x,t)=0
\label{SG-1.1}
\end{equation}
appeared in the \Roman{nineteen} century Geometry in the light-cone
representation

\begin{equation}
u_{\xi\eta}=4\sin u, \ \ \ u=u(\xi,\eta),
\label{SG-1.2}
\end{equation}
where
\begin{equation}
x=2(\xi+\eta), \ \ t=2(\xi-\eta), \ \
\partial_\xi=2\partial_x+2\partial_t, \ \
\partial_\eta=2\partial_x-2\partial_t.
\label{SG-1.3}
\end{equation}

A broad family of the ``soliton-type solutions'' was found already
in  XIX century: it was based on the discovery by Bianchi and
S.Lie
 the substitution
called later  the ``Backlund Transformation''.

The analogy of SG with KdV (i.e. association with the so-called
``Inverse Scattering Transform'') was found in the early 1970s by
G.Lamb (see \cite{Lamb}). The modern approach to the study  SG
equation was developed in the work \cite{AKNS} in 1974. It is based on
the representation of the SG equation as a consistency condition for the following
linear systems rationally dependent on the parameter $\lambda$:
\begin{equation}
\Psi_\xi=U\Psi, \ \ \Psi_\eta=V\Psi,
\label{SG-1.4}
\end{equation}
where
\begin{equation}
\Psi=\Psi(\lambda,\xi,\eta)=\left(\begin{array}{c} \psi_1(\lambda,\xi,\eta) \\
\psi_2(\lambda,\xi,\eta) \end{array}\right),
\label{SG-1.5}
\end{equation}

\begin{equation}
U=U(\lambda,\xi,\eta)=\left[\begin{array}{cc}
\frac{iu_\xi}{2} & 1 \\ -\lambda  & -\frac{iu_\xi}{2}
\end{array}\right], \ \
V=V(\lambda,\xi,\eta)=\left[\begin{array}{cc}
0 & -\frac{1}{\lambda}e^{iu} \\ e^{-iu} & 0.
\end{array}\right]
\label{SG-1.6}
\end{equation}

In the variables  $(x,t)$ we have:
\begin{equation}
\Psi_x=\frac14 (U+V)\Psi, \ \ \Psi_t=\frac14(U-V)\Psi,
\label{SG-1.7}
\end{equation}

where

\begin{equation}
U=U(\lambda,x,t)=\left[\begin{array}{cc}
i(u_x+u_t) & 1 \\ -\lambda  & -i(u_x+u_t)
\end{array}\right],
\label{SG-1.8}
\end{equation}

\begin{equation}
V=V(\lambda,x,t)=\left[\begin{array}{cc}
0 & -\frac{1}{\lambda}e^{iu} \\ e^{-iu} & 0.
\end{array}\right].
\label{SG-1.9}
\end{equation}

\begin{definition} We call the compatibility condition for pair of linear systems
like above (see (\ref{SG-1.4}), (\ref{SG-1.7}) a (rationally) {\bf  $\lambda$-dependent
Zero Curvature representation}. It can be written in algebraic form
$$
[\partial_{\xi}-U(\lambda),\partial_\eta-V(\lambda)]=0.
$$
\end{definition}

 Let us mention that  the standard Lax representation
 for the KdV and all higher KdV equations
 $dL/dt_n=[A_n,L]$ based on the $\lambda$-independent linear
  differential operators
 $L,A_n$, has been transformed into the (polynomially) $\lambda$-dependent
 zero-curvature representation in the work \cite{NOV} for the needs
 of the periodic problem (to study the finite-gap solutions invented in
this work). Very interesting elliptic zero curvature representations
appeared later. Until now we do not know interesting effectively written
zero-curvature representations with more complicated $\lambda$-dependence.

\begin{definition} A SG solution $u(x,t)$ is called {\bf periodic}  in $x$
if the
quantity $e^{iu}$ is periodic with period $T$ in the real variable $x\in R$ .
\end{definition}

\begin{definition} The quantity $n=[u(x+T)-u(x)]/2\pi$, where $T$
is the period is called {\bf topological charge.} The ratio $\bar
n=n/T$ is called {\bf density of topological charge.}
\end{definition}

The following statement is well-known in the theory of functions:

\begin{lemma}
\label{SG-L0.1}
 Let $u(\vec X)$, $X\in{\Bbb R}^n$ be a smooth function in ${\Bbb R}^n$
 such, that $\exp(iu(\vec X))$ is single-valued on the torus
 ${\Bbb R}^n/{\Bbb Z}^n$, i.e. $\exp(iu(\vec X+\vec N))=\exp(iu(\vec X))$
 for any integer vector $\vec N$. Denote by $u(x)$ restriction of
 $u(\vec X)$ to the strait line $\vec X=\vec X_0 +x\cdot\vec v$.
 Then the density of topological charge $\bar n=\lim\limits_{T\rightarrow\infty}
 [u(x+T)-u(x)]/2\pi T$ is well-defined; it does not depend on the
 point $\vec X_0$ and can be expressed by the following formula:
\begin{equation}
\bar n=\sum\limits_{k=1}^n n_k v^k,
\label{SG-1.10}
\end{equation}
where $\vec v=(v^1,v^2,\ldots,v^n)$, and $n_k$ are topological
charges along the basic cycles ${\cal A}_k,k=1,\ldots,n$:
\begin{equation}
u(X^1,X^2,\ldots,X^k+1,\ldots,X^n)-u(X^1,X^2,\ldots,X^k,\ldots,X^n)=2\pi n_k.
\label{SG-1.11}
\end{equation}
\end{lemma}
The {\bf Topological Charge along the cycle} ${\cal A}=\sum
l_k{\cal A}_k\in H_1(T^n,Z)$ is equal to the linear combination
$n({\cal A})=\sum l_kn_k$ by definition.

\subsection{Algebro-geometric solutions: The complex theory.}

Let us recall the construction of general complex ``finite-gap''
SG solutions. They were constructed almost immediately after the
finite-gap KdV theory (see \cite{KK,IK}).

Take the following ``spectral data'' similar to KdV:
\begin{enumerate}
\item A nonsingular hyperelliptic Riemann surface
    $\Gamma$ $[\mu^2=R(\lambda)]$, where
      $R(\lambda)=\prod\limits_{k=0}^{2g}(\lambda-E_k)$, such that $E_0=0$
      and $E_i\neq E_j, i,j=0,1,\ldots,2g$.
      It has exactly $2g+2$ branching points $E_0=0$, $E_1$, \ldots,
      $E_{2g}$, $\infty$; the genus of $\Gamma$ is equal to $g$. A point
      $\gamma\in\Gamma$ is by definition a pair of complex numbers
      $\gamma=(\lambda,\mu)$ such, that $\mu^2=R(\lambda)$.

\item A divisor $D$ of degree $g$, i.e. set (or formal sum) of $g$ points
$D=\gamma_1+\ldots+\gamma_g$.
\end{enumerate}

 For generic data $\Gamma$, $D$ there exists an
unique two-component ``Baker-Akhiezer''
 vector-function $\Psi(\gamma,\xi,\eta)$ such that
\begin{enumerate}
\item For fixed $(\xi,\eta)$ the function $\Psi(\gamma,\xi,\eta)$ is meromorphic
    in the variable
      $\gamma\in\Gamma$ outside the points $0$, $\infty$ and has at
      most 1-st order poles at the divisor points $\gamma_k$, $k=1,\ldots,g$.
\item  $\Psi(\gamma,\xi,\eta)$ has essential singularities at the points
      $0$, $\infty$ with the following asymptotics:
\begin{equation}
\Psi(\gamma,\xi,\eta)=\left(\begin{array}{c}
1+o(1) \\ i\sqrt{\lambda}+O(1)
\end{array} \right)e^{i\sqrt{\lambda}\xi} \ \ \mbox{as} \ \ \lambda\rightarrow\infty,
\label{SG-1.12}
\end{equation}
\begin{equation}
\Psi(\gamma,\xi,\eta)=\left(\begin{array}{c}
\phi_1(\xi,\eta)+o(1) \\ i \sqrt{\lambda} \phi_2(\xi,\eta)+O(\lambda)
\end{array} \right)e^{-\frac{i}{\sqrt{\lambda}}\eta} \ \ \mbox{as} \ \
\lambda\rightarrow 0,
\label{SG-1.13}
\end{equation}
where $\phi_1(\xi,\eta)$, $\phi_2(\xi,\eta)$ are some functions of
the variables $\xi,\eta$ .
\end{enumerate}

Let us point out that the divisor of zeroes of the first component
$\psi_1$ we denote $D(\xi,\eta)=\sum_j\gamma_j(\xi,\eta)$. For
$\xi=0$, $\eta=0$ we have $D(0,0)=D$. The equations for this
divisor in the variables $\xi$, $\eta$ we call Dubrovin equations.
The are important for our work (see below). The vector-function
$\Psi(\gamma,\xi,\eta)$  satisfies to the zero-curvature equations
(\ref{SG-1.4}) with  potential given by the formula
\begin{equation}
u(\xi,\eta)=i\ln \frac{\phi_2(\xi,\eta)}{\phi_1(\xi,\eta)}.
\label{SG-1.14}
\end{equation}
Therefore the function $u(\xi,\eta)$ solves the SG equation
(\ref{SG-1.2}).

Both functions $\Psi$ and $u$ can be expressed through the
$\Theta$-functions associated with  Riemann surface $\Gamma$. It
is easier to express the solution $u$ through the
$\Theta$-functions than $\Psi$. Let us mention that the first
$\Theta$-functional formula for the functions like $\Psi$ was
obtained  by A.Its  in the case of KdV; it was found  later than
the Its-Matveev formula for the potential and KdV solution
$u(x,t)$--see the \cite{DMN}, Appendix 1. In the higher rank
problems there exist sometimes explicit formulas for the functions
like $u$, but $\Psi$ cannot be found explicitly
(see\cite{KricNov}). As a second remark let us point out that the
first component $\psi_1$ of the Baker-Akhiezer vector-function
$\Psi$ is a partial case of the general scalar two-point
Baker-Akhiezer functions satisfying to the second order linear
Schr\"odinger equation $L_1\psi_1=0$ invented in the work
\cite{DKN}. Here we have
$$L_1=\frac{\partial^2}{\partial \xi\partial\eta} +A(\xi,\eta)\frac{
\partial }{\partial \eta} +W(\xi,\eta),$$ $-A=\partial_\xi\log\phi_1$.
In this special case our surface $\Gamma$ is hyperelliptic,  the
selected points coincide with $0,\infty$ and corresponding local
parameters are chosen as $\sqrt{\lambda},1/\sqrt{\lambda}$. The
second function $\psi_2$ satisfies to similar equation
$L_2\psi_2=0$ where the operator $L_2$ is obtained from $L_1$ by
the  so-called Laplace Transformations  and vice versa. It is
known that cyclic Laplace Chains of the length 2 lead to the SG
equation in the general complex case (see in the work \cite{NV2}).

For the generic ``spectral data'' $\Gamma$, $D$  corresponding SG
solutions are  complex and may have singularities for real $x$.
The function $\exp\{iu\}$ can be naturally considered as a certain
meromorphic function on the complex Jacoby torus
$T^{2g}=J(\Gamma)$ restricted to a linear subspace.

For the effective reconstruction of the function $u(\xi,\eta)$ we use
the formula

\begin{equation}
e^{iu(\xi,\eta)}=\frac{\prod\limits_{j=0}^{g}(-\lambda_j(\xi,\eta))}
{\sqrt{\prod\limits_{j=1}^{2g} E_j}} \label{SG-13}
\end{equation}

Dubrovin  equations in the variables $\xi,\eta$ for the divisor $D(\xi,\eta)$
of zeroes of the first function $\psi_1$,
where $D(\xi,\eta)=\gamma_1(\xi,\eta)+\ldots+\gamma_g(\xi,\eta)$ and
$\gamma_j(\xi,\eta)=(\lambda_j(\xi,\eta),\mu_j(\xi,\eta))$,
have the form:

\begin{equation}
\left\{
\begin{array}{l}
\frac{\displaystyle \partial\lambda_k(\xi,\eta)}{\displaystyle \partial \xi}=
\displaystyle -2i\ \frac{\displaystyle \mu_k(\xi,\eta)}
{\displaystyle \prod\limits_{j\ne k}(\lambda_k(\xi,\eta)-\lambda_j(\xi,\eta))}\\ \\
\frac{\displaystyle \partial\mu_k(\xi,\eta)}{\displaystyle \partial \xi}=
\displaystyle -i\ \frac{\displaystyle R'(\lambda_k(\xi,\eta))}
{\displaystyle \prod\limits_{j\ne k}(\lambda_k(\xi,\eta)-\lambda_j(\xi,\eta))}\\ \\
\frac{\displaystyle \partial u(\xi,\eta)}{\displaystyle \partial \xi}=
\displaystyle -2\ \sum\limits_{k}\frac{\displaystyle \mu_k(\xi,\eta)}
{\displaystyle \lambda_k(\xi,\eta)\prod\limits_{j\ne k}
(\lambda_k(\xi,\eta)-\lambda_j(\xi,\eta))},
\end{array}\right.
\label{SG-14}
\end{equation}
where

\begin{equation}
R'(\lambda)=\frac{\partial R(\lambda)}{\partial\lambda}
\label{SG-15}
\end{equation}

\begin{equation}
\left\{
\begin{array}{l}
\frac{\displaystyle \partial\lambda_k(\xi,\eta)}{\displaystyle \partial \eta}=
\displaystyle 2i\ \frac{\displaystyle \mu_k(\xi,\eta) e^{iu(\xi,\eta)}}
{\displaystyle \lambda_k(\xi,\eta)\prod\limits_{j\ne k}
(\lambda_k(\xi,\eta)-\lambda_j(\xi,\eta))}\\ \\
\frac{\displaystyle \partial\mu_k(\xi,\eta)}{\displaystyle \partial \eta}=
\displaystyle i\ \frac{\displaystyle R'(\lambda_k(\xi,\eta)) e^{iu(\xi,\eta)}}
{\displaystyle \lambda_k(\xi,\eta)\prod\limits_{j\ne k}
(\lambda_k(\xi,\eta)-\lambda_j(\xi,\eta))}\\ \\
\frac{\displaystyle \partial u(\xi,\eta)}{\displaystyle \partial \eta}=
\displaystyle 2\ \sum\limits_{k}\frac{\displaystyle \mu_k(\xi,\eta) e^{iu(\xi,\eta)}}
{\displaystyle  \lambda_k(\xi,\eta)^2\prod\limits_{j\ne k}
(\lambda_k(\xi,\eta)-\lambda_j(\xi,\eta))},
\end{array}\right.
\label{SG-16}
\end{equation}
$e^{iu}$ is defined by (\ref{SG-13}).

\subsection{Real finite-gap SG solutions: known results.}
\label{SG-Sect2.3}

It turns out that the simplest problem here is to find the class
of Riemann surfaces  corresponding to the real solutions. This
class was found in the work \cite{IK}:

 The set of branching points of the Riemann surface $\Gamma$ should contain
  real points or  complex conjugate pairs of points only.
       All real branching points should be nonpositive.
       Therefore we may assume, that the first $2m+1$ of them
       $E_0=0$, $E_1$, $E_2$,
      \ldots, $E_{2m}$ are real and $0>E_1>E_2>\ldots>E_{2m}$. For
      $k>m$ we have
      $E_{2k}=\overline{E_{2k-1}}$, $\Im E_{2k}\ne 0$.
      Therefore our Riemann surface (algebraic curve) $\Gamma$ is real
       and admits a ${\Bbb Z}_2\times{\Bbb Z}_2$-group  of involutions
       generated by
       the standard holomorphic involution $\sigma:\Gamma\rightarrow\Gamma$
       such that
      \begin{equation}
       \sigma(\lambda,\mu)=(\lambda,-\mu),
    \label{SG-1.15}
      \end{equation}
      and by the antiholomorphic involution $\tau$ such that
      $\tau_1=\sigma\tau=\tau\sigma$:
      \begin{equation}
       \tau(\lambda,\mu)=(\bar\lambda,\bar\mu).
    \label{SG-1.16}
      \end{equation}

Which divisors on these surfaces generate real SG solutions?

They were found in the work \cite{Cher}.  Let us formulate this
result.

Take a real Riemann surface of the class described above. Try to
construct  a meromophic differential $\Omega$ with two simple
poles located in the points $0,\infty$ and with  $2g$ zeroes
located in the points $D+\tau D$ where
$D=\gamma_1+\ldots,\gamma_g$.

\begin{definition}. We call the divisor $D$ {\it admissible} if and
only if such differential $\Omega$ exists.
\end{definition}

The admissible divisors lead to the real nonsingular solutions of
the SG equation. According to the result of \cite{Cher} any real
solution is nonsingular. So the property of divisor to be
admissible  is sufficient for the solution to be real and smooth
quasiperiodic function for the nonsingular real algebraic curve
$\Gamma$.

We shall explain below (see Appendix 1) that this requirement is
also necessary for the periodic solution to be real nonsingular.

Let us demonstrate here a simple proof of the statement:

 The potential
$u(\xi,\eta)$ is real if the condition formulated above is
satisfied

Proof. Consider the following 1-forms:

\begin{equation}
\label{SG-1.17}
\Omega_1=\psi_1(\gamma,\xi,\eta)\overline{\psi_1(\tau\gamma,\xi,\eta)}\Omega,\ \ \
\Omega_2=\frac{-1}{\lambda}\psi_2(\gamma,\xi,\eta)\overline{\psi_2(\tau\gamma,\xi,\eta)}
\Omega.
\end{equation}
We may assume, that the residues of the form  $\Omega$ are equal
to the number $\pm 1$ at the points $0$, $\infty$ respectively.
Then the residues of the differentials $\Omega_1$, $\Omega_2$ at
the point $\infty$ are equal  to the number $-1$. Calculating the
residues at the point $0$, we get:
\begin{equation}
\label{SG-1.18}
1=\phi_1(\xi,\eta)\overline{\phi_1(\xi,\eta)}=
\phi_2(\xi,\eta)\overline{\phi_2(\xi,\eta)}.
\end{equation}
Therefore we have
\begin{equation}
\label{SG-1.19}
\left|\frac{\phi_2(\xi,\eta)}{\phi_1(\xi,\eta)}\right|=1
\end{equation}
and the potential $u(\xi,\eta)$ in (\ref{SG-1.14}) is real. Our
statement is proved.

The form $\Omega_1$ defined above has simple poles at the points $0$, $\infty$
and zeroes at the points $\gamma_1(\xi,\eta)$, \ldots, $\gamma_g(\xi,\eta)$,
$\tau\gamma_1(\xi,\eta)$, \ldots, $\tau\gamma_g(\xi,\eta)$. Therefore the divisor
of zeroes of $\psi_1$ $D(\xi,\eta)=\gamma_1(\xi,\eta)+\ldots+\gamma_g(\xi,\eta)$
is admissible for all $\xi,\eta$.

In the work \cite{Cher} the number of real components was found:
it is equal to  $2^m$ where $2m$ is a number of the strictly
negative branching points. No effective description of these
components was presented in this work.
 In the later works \cite{Dubr-Nat},\cite{Erc-Forest} all real
 components were found and described in terms of Jacobian variety
 associated with
 $\Theta$-functions. It became clear that this description is very hard to use
 for the studying simple fundamental properties of solutions.
 For example, this description do not imply any
 formula for the topological charge. A most recent survey of this
 subject can be found in \cite{Nat}.

 In the joint work \cite{DN}
 an attempt has been made to describe real components in homological terms
 involving the cycles $[\gamma_j]\in H_1(\Gamma \backslash (0\bigcup \infty)
 ,Z)$ covered by the moving zeroes $\gamma_j(x)$ of the
 Baker-Akhiezer function $\psi_1$ on the Riemann surface $\Gamma$.
 In the special parts of the space of parameters this idea led to
 the interesting formulas for the topological charge and action
 variables in \cite{DN}. However, an attempt to extend it to the
 whole phase space failed as it mentioned in \cite{Nov}.
 In the present work we are going to prove formula for the
 topological charge proposed in \cite{DN,Nov}.

\section{Effective description of the real tori.
Admissible polynomials}

Let us consider only real Riemann surfaces of the type already
described above. To construct a real solution to the SG equation
we need to choose a meromorphic differential $\Omega$ described in
the previous paragraph.

\begin{lemma}
\label{lemma1}

The differential $\Omega$ can be expressed in the following form
\begin{equation}
\label{SG-2.1}
\Omega=\left(1-\frac{\lambda P_{g-1}(\lambda)}{R(\lambda)^{1/2}}\right)
\frac{d\lambda}{2\lambda},
\end{equation}
\end{lemma}
Here $P_{g-1}(\lambda)$ is a polynomial of degree at most $g-1$
with real coefficients such, that the system
\begin{equation}
\label{SG-2.2}
\left\{
\begin{array}{c}
\mu=\lambda P_{g-1}(\lambda) \\
R(\lambda)=\mu^2,
\end{array}\right.
\end{equation}
has  solutions exactly in the points
$(0,0),\gamma_1,\ldots,\gamma_g,
\tau\gamma_1,\ldots,\tau\gamma_g$.

As it was already mentioned, the proof of this lemma immediately
follows from the general classification of the meromorphic
differentials on the surface $\Gamma$. We simply represented it in
this standard form.

\begin{definition}. We call a real  polynomial $P_{g-1}$ of the
degree at most $g-1$ admissible if there exists a divisor
$D=\gamma_1+\ldots=\gamma_g$ such the system (\ref{SG-2.2}) has
solutions in the points $(0,0)$ and $D+\tau D$. In particular, any
real solution of this system (\ref{SG-2.2}) except $(0,0)$ should
have even multiplicity.
\end{definition}

Admissible polynomials $P_{g-1}(\lambda)$ can be characterized in
the following way:

Consider a pair of functions of the real variable $\lambda$:
\begin{equation}
f_{\pm}(\lambda)=\pm\frac{\sqrt{R(\lambda)}}{\lambda}.
\label{SG-2.3}
\end{equation}
We assume, that the functions $f_{\pm}(\lambda)$ are defined only
for such $\lambda\ne0$ that $R(\lambda)\ge0$, i.e. these functions
are real-valued. Let us draw the graphs of these functions
$y=f_+(\lambda)$, $y=f_-(\lambda)$, and  fill in the following
domains by the black color:
\begin{equation}
\begin{array}{l}
\lambda<0, y^2 < \frac{R(\lambda)}{\lambda^2}, \\
\lambda>0, y^2 > \frac{R(\lambda)}{\lambda^2}.
\end{array}
\label{SG-2.4}
\end{equation}

\vspace{1cm}

\begin{center}
\mbox{\epsfxsize=10cm \epsffile{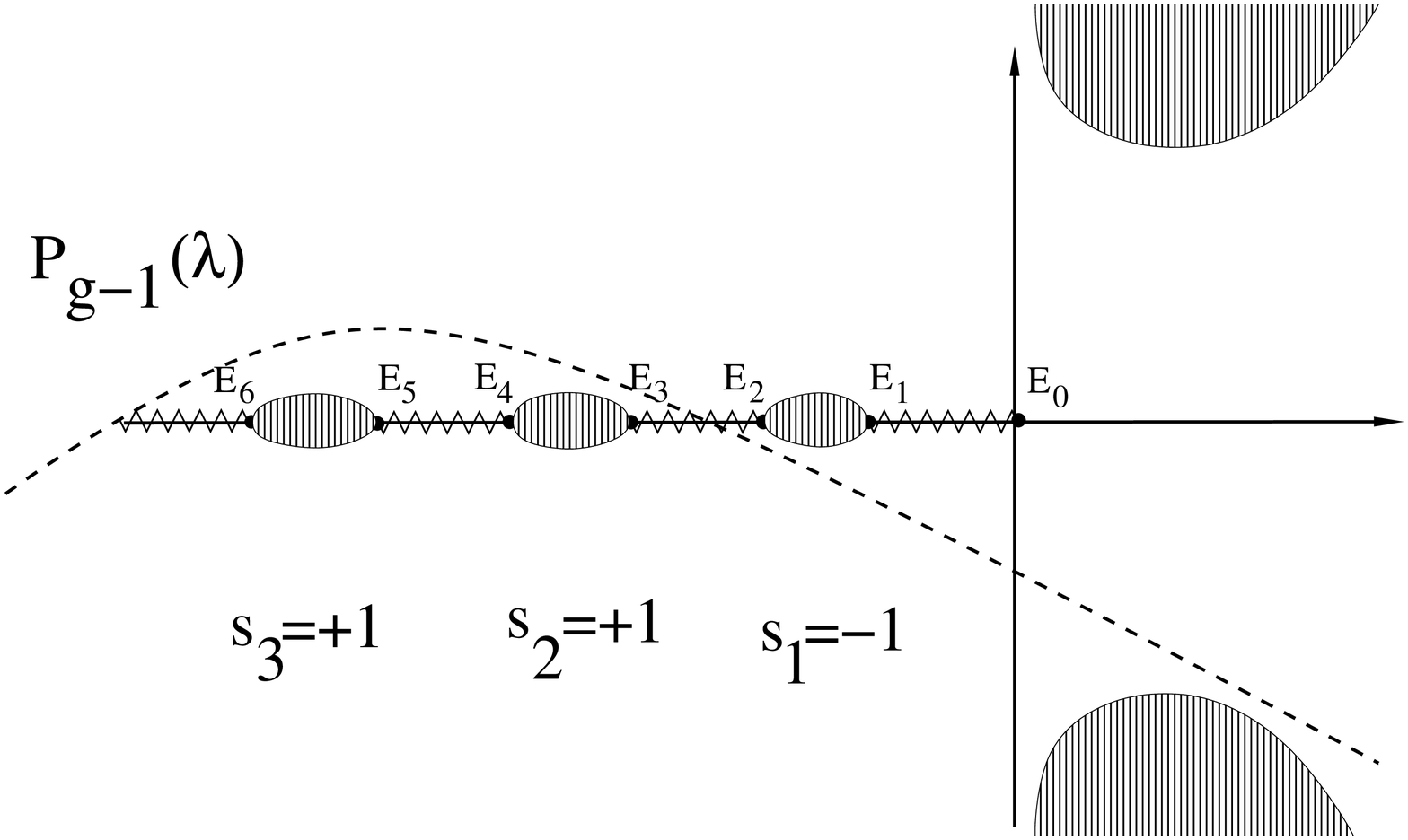}}

Fig 1.
\end{center}

\begin{lemma}
\label{SG-L1} The polynomial $P_{g-1}(\lambda)$ is admissible if
and only if the graph of $P_{g-1}(\lambda)$ does not cross the
black open domains (but it can touch their boundaries).
\end{lemma}

We associate an admissible polynomial with an admissible divisor.
Denote this map by $\Pi:D\rightarrow P_{g-l}(\lambda)$. The
inverse map $\Pi^{-1}$ is multivalued. If $P_{g-1}(\lambda)$ is an
admissible polynomial, then the equation (\ref{SG-2.2}) has $2g+1$
roots. One of them is the point $(0,0)$, other $2g$ roots form $g$
pairs. To define an admissible divisor we have to choose one point
from each pair: therefore we have at most $2^g$ possibilities
depending on the number of real points. In the generic case we
have no real roots, so the number is equal to $2^g$ in this case.

It follows immediately from Lemma~\ref{SG-L1} that the set of all
admissible divisors is compact.

We already know that the set of all admissible divisors consists
of $2^m$ connected components. But our knowledge of this fact is
noneffective yet: it simply follows from the number of real
components found by the different authors many years ago (see the
previous section). We are going to describe real components
effectively in terms of the admissible polynomials.

Let us associate with each admissible polynomial a {\bf
Topological Type}, i.e. a collection of $m$ numbers $s_k=\pm1$,
$k=1,\ldots,m$ defined by the following rule:

$s_k=1$ if $P_{g-1}(\lambda) \ge f_+(\lambda)$ as
$E_{2k}\le\lambda\le E_{2k-1}$,

$s_k=-1$ if $P_{g-1}(\lambda) \le f_-(\lambda)$ as
$E_{2k}\le\lambda\le E_{2k-1}$.

\begin{lemma}
\label{SG-L1.1} Each topological type is presented by the convex
subset in the space of polynomials $P_{g-1}$.
\end{lemma}

Proof. Each topological type is defined by the system of linear
inequalities as it follows from the definition above. Lemma is
proved.

\begin{lemma}
\label{SG-L1.2} Each of these $2^m$ connected components is
non-empty and depends continuously on the branching points
$E_1$,\ldots,$E_{2g}$.
\end{lemma}

It is sufficient for the proof to point out, that for any
collection of $m$ numbers $s_k=\pm1$, $k=1,\ldots,m$ the
polynomial

\begin{equation}
P_{g-1}(\lambda)=\sqrt{-E_1-E_{2m}}(-1)^{s_1}
\prod\limits_{k=2}^m (\sqrt{E_{2k-2} E_{2k-1}}-(-1)^{s_k
s_{k-1}}\lambda)
\prod\limits_{k=m+1}^g(\sqrt{E_{2k-1}E_{2k}}-\lambda)
\label{SG-2.5}
\end{equation}
is admissible and belongs to the appropriate connected component.

The properties of the polynomial (\ref{SG-2.5}) can be easily
derived from the following elementary inequalities.

\label{SG-L1.3} Let $a$, $b$ be a pair of negative real numbers,
$a<b$. Then:
\begin{equation}
\label{SG-2.5.1}
\begin{array}{l}
\left| \sqrt{\frac{(x-a)(x-b)}{x}} \right|
> \sqrt{-a-b} \ \ \mbox{as} \ \ x>0, \\
\left| \sqrt{\frac{(x-a)(x-b)}{x}} \right| < \sqrt{-a-b} \ \
\mbox{as} \ \ a<x<b.
\end{array}
\end{equation}

\label{SG-L1.4} Let $a$, $b$ be a pair of negative real numbers,
$a<b$. Then
\begin{equation}
\label{SG-2.5.2}
\begin{array}{l}
\left| \sqrt{(x-a)(x-b)} \right|
> \left|(\sqrt{ab}\pm x) \right| \ \ \mbox{as} \ \ x>0, \\
\left| \sqrt{(x-a)(x-b)} \right|
< \left|(\sqrt{ab}\pm x) \right| \ \  \mbox{as} \ \ b<x<0, \\
\left| \sqrt{(x-a)(x-b)} \right| < \left|(\sqrt{ab}\pm x) \right|
\ \  \mbox{as} \ \ x<a.
\end{array}
\end{equation}

\label{SG-L1.5} Let $a$, $b$ be a complex number with non-zero
imaginary part, $f(x)=\sqrt{(x-a)(x-\bar a)}$. Then
\begin{equation}
\label{SG-2.5.3}
\begin{array}{l}
\left| \sqrt{(x-a)(x-\bar a)} \right| >
\left|(\sqrt{a\bar a}- x) \right|\ \ \mbox{as} \ \ x>0, \\
\left| \sqrt{(x-a)(x-\bar a)} \right| < \left|(\sqrt{a\bar a}- x)
\right| \ \  \mbox{as} \ \ x<0.
\end{array}
\end{equation}

Combining the inequalities  (\ref{SG-2.5.1})-(\ref{SG-2.5.3}) we are coming
to the conclusion:

\label{SG-L1.6} For the polynomial $P_{g-1}(\lambda)$ defined by
the formula (\ref{SG-2.5}) we have the following inequalities:

\begin{equation}
\label{SG-2.6}
\begin{array}{l}
\left|\frac{\sqrt{R(\lambda)}}{\lambda}\right| >
\left|P_{g-1}(\lambda) \right|\ \ \mbox{as} \ \ x>0, \\
 \left|\frac{\sqrt{R(\lambda)}}{\lambda}\right| <
\left|P_{g-1}(\lambda) \right| \ \  \mbox{as}  \ \ x<0 \ \
\mbox{and the left-hand side is real}.
\end{array}
\end{equation}

Therefore the polynomial $P_{g-1}(\lambda)$ is admissible. Lemma is
proved.

The following property of admissible divisors is essential:

\begin{lemma}
\label{SG-L2} Projections of the points of admissible divisors to
the $\lambda$-plane could not lie in the segments $[E_1,E_0]$,
$[E_3,E_2]$,\ldots,$(-\infty,E_{2m}]$. Moreover, it is possible to
choose an open neighborhood of these segments such, that the
projections of these points could not lie in it.
\end{lemma}

{\it Proof of Lemma~\ref{SG-L2}}. If $(\lambda_k.\mu_k)$ is a
point of admissible divisor such, that
\begin{enumerate}
\item $\lambda\in{\Bbb R}$.
\item $E_{2j}<\lambda_k<E_{2j-1}$, $1\le j\le m$,
\end{enumerate}
then $R(\lambda_k)<0$ and $\mu_k$ is a pure imaginary number. But
from the description of admissible divisors we have $\mu_k=\lambda
P_{g-1}(\lambda_k)$ and $\mu_k$ should be real. We have obtained a
contradiction.

Assume now, that $\lambda_k=E_j$ where $E_j$ is one of real
branching points, $j\ne 0$. Then $\mu_k=0$ and
$P_{g-1}(\lambda_k)=0$. In the neighborhood of $E_j$ the functions
$f_{\pm}(\lambda)$ behaves like $\sqrt{c(\lambda-E_j)}$; therefore
the graph of $P_{g-1}(\lambda)$ intersects the corresponding black
domain. We have obtained a contradiction.

The existence of such a neighborhood follows from the compactness
of the set of admissible divisor.

Lemma~\ref{SG-L2} is proved.

\begin{lemma}
\label{SG-L2.1} Let projection of a point
$\gamma_s=(\lambda_s,\mu_s)$ of an admissible divisor to the
$\lambda$-plane lie in the segments $[E_{2k},E_{2k-1}]$; then we
have either $\mu_s<0$ for the case $s_k=1$ or $\mu_s>0$ if
$s_k=-1$.
\end{lemma}

{\it Proof of Lemma~\ref{SG-L2.1}}.
By definition $s_k$ is the sign of the polynomial $P_{g-1}(\lambda_s)$ at this
interval. Taking into account that $\mu_s=\lambda_s P_{g-1}(\lambda_s)$ and
$\lambda_s<0$ we complete the proof.

\section{Topological charge.}

We call the spectral curve $\Gamma$ {\bf stable} if all branching
point are real, i.e. $m=g$. It is convenient for the stable $\Gamma$
to choose a real canonical basis of cycles in the following form.

\begin{center}
\mbox{\epsfxsize=10cm \epsffile{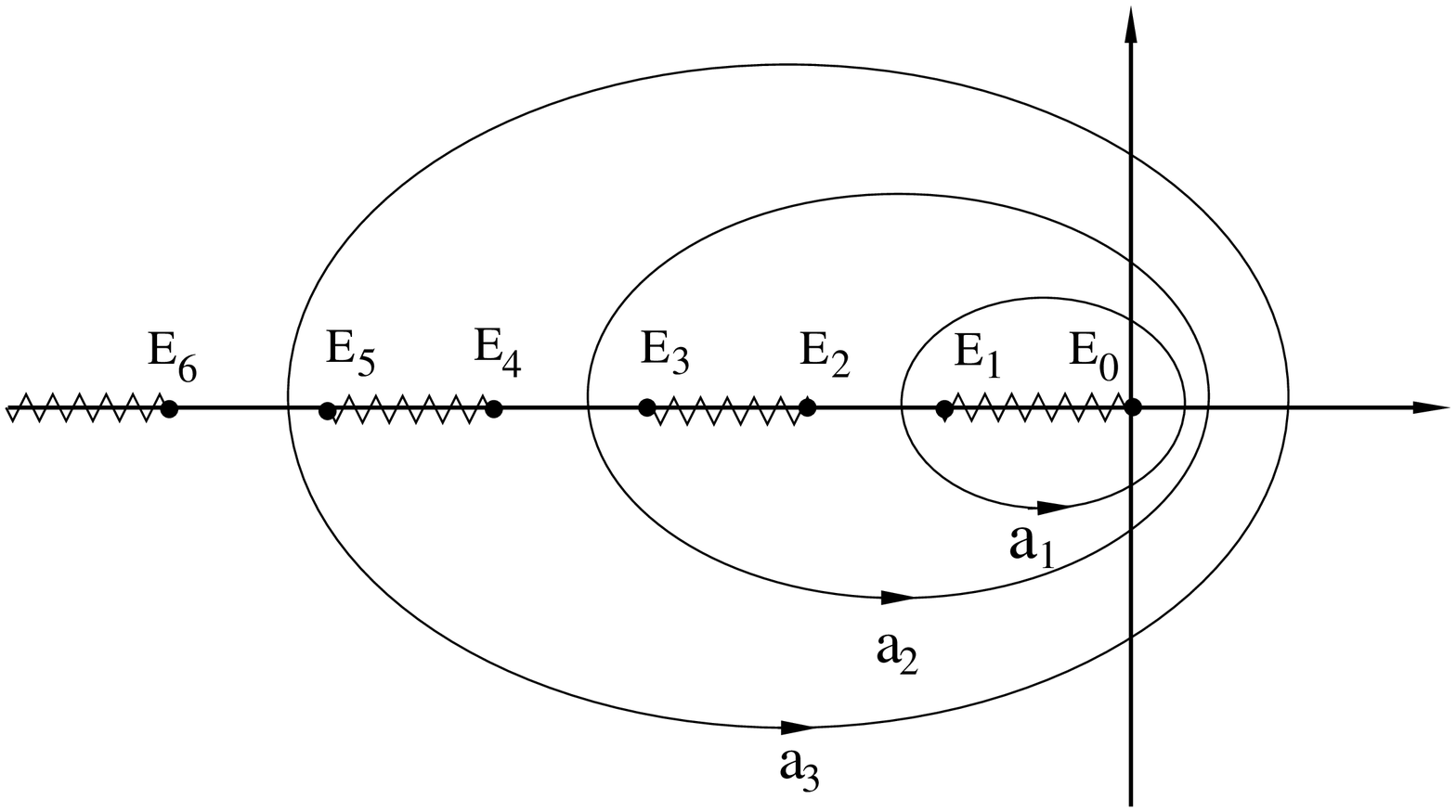}}

Fig 2.
\end{center}

Let us draw the following system of cuts $[E_1,E_0]$,
$[E_3,E_2]$,\ldots, $(-\infty,E_{2g}]$. After making these cuts we
get 2 sheets. It is specific property of the stable case.
Let us denote the sheet containing the line
$\lambda\in{\Bbb R}$, $\lambda>0$ $\mu>0$ by $G_+$ and the second
sheet by $G_-$ (see Fig~2).

For the general real case we have exactly $2m$ negative branching points
$E_1,\ldots,E_{2m}$ and $g-m$ complex adjoint pairs $E_{2j-1},E_{2j}=\bar E_{2j-1}$,
$j=m+1,\ldots,g$. The first $m$ $a$-cycles should be chosen as in the stable case.
The last $g-m$ $a$-cycles should be chosen as coverings on $\Gamma$ over the
path on the $\lambda$-plane connecting the points $E_{2j-1}$ and $E_{2j}$ for
$j>m$. All these path should not meet each other. All of them should cross
positive part of the line $\lambda>0$ in one point $\kappa_j$ (i.e. in two
points of Riemann surface) such that $\kappa_{m+1}<\kappa_(m+2)<\ldots < \kappa_g$.
This basis depends on the order of the complex conjugate pairs only (see Fig~3).

\begin{center}
\mbox{\epsfxsize=10cm \epsffile{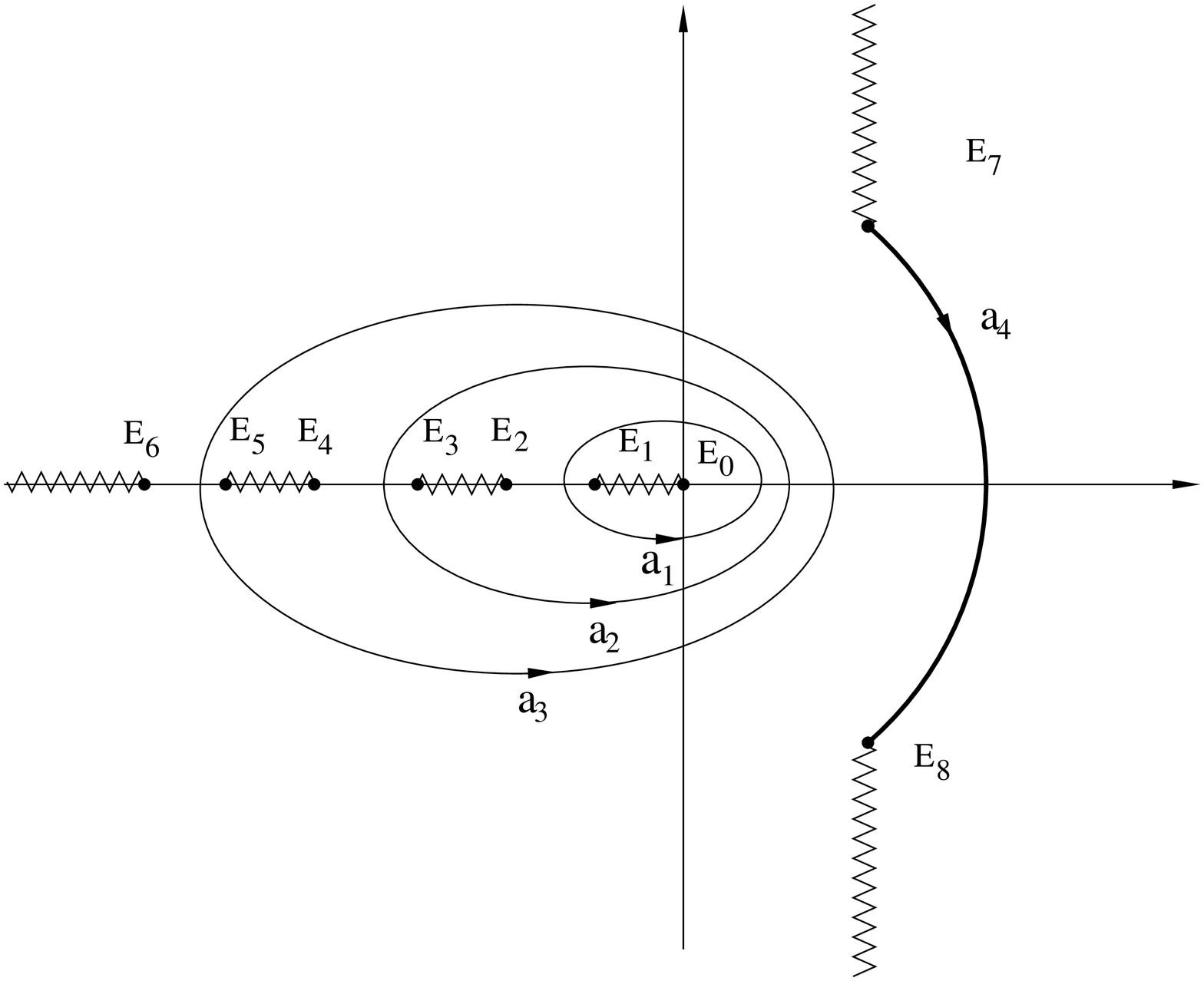}}

Fig 3.
\end{center}

For the complex conjugate pairs of branching points
$E_{2j},E_{2j-1}, j>m,$ we make the cuts connecting each of these
points with $i\infty$ or $-i\infty$ such that the cuts do not
intersect the real line. As before we denote the sheet containing
the line $\lambda\in{\Bbb R}$, $\lambda>0$ $\mu>0$ by $G_+$ and
the second sheet by $G_-$.

Denote by  $\omega^1$,\ldots,$\omega^{g-1}$ the canonical basis of
holomorphic differentials in $\Gamma$,
\begin{equation}
\label{SG-3.1} \omega^k=i\ \frac{\sum\limits_{j=0}^{g-1}
D^k_j\lambda^j}{\mu}\ d\lambda, \ \ D^k_j\in{\Bbb R}
\end{equation}

\begin{theorem}
For any real solution the density of topological charge
along the variable $x$ is given by the formula
\begin{equation}
\label{SG-3.2}
\bar n=\frac12 \ \sum\limits_{k=1}^{m} (-1)^{k-1} s_k
\left(D^k_{g-1}+
\frac{D^k_{0}}{\sqrt{\prod\limits_{j=1}^{2g}E_j}}\right).
\end{equation}
\end{theorem}

{\bf Proof}. It follows from the $\Theta$-functional formulas that
the function $\exp\{iu(x,t)\}$ is a regular function on a real $g$
dimensional torus restricted to a two-dimensional linear subspace.
Each such torus corresponds to the fixed topological type; it is a
real component of the Jacoby torus $T^{2g}=J(\Gamma)$ with respect
to an antiholomorphic involution. Therefore we can apply
Lemma~\ref{SG-L0.1}.

Our calculation consists of two steps. At the first step we calculate the
coefficients of the vector $\vec v$, and at the second step we calculate
the topological charges along the basic cycles.

The first step is rather standard (see, for example, \cite{DMN}).
Dubrovin equations (\ref{SG-14}), \ref{SG-16}) can be written in the
following form

\begin{equation}
\frac{\partial\lambda_k(\xi,\eta)}{\displaystyle \partial \xi}=
-\mbox{res}_{\gamma=\gamma_k(\xi,\eta)}
(\partial_\xi \log \psi_1(\gamma,\xi,\eta))
d\lambda,
\label{SG-3.2.1}
\end{equation}

\begin{equation}
\frac{\partial\lambda_k(\xi,\eta)}{\displaystyle \partial \eta}=
-\mbox{res}_{\gamma=\gamma_k(\xi,\eta)}
(\partial_\eta \log \psi_1(\gamma,\xi,\eta))
d\lambda,
\label{SG-3.2.2}
\end{equation}

From (\ref{SG-3.2.1}), (\ref{SG-3.2.2}) we obtain following formulas
for the $\xi$, $\eta$ derivatives of the Abel transform of the divisor:

\begin{equation}
\label{SG-3.2.3}
\frac{\partial A^k(\vec\gamma)}{\partial \xi}=
-\sum_k \mbox{res}_{\gamma=\gamma_k(\xi,\eta)} \omega^k
\partial_\xi \log \psi_1(\gamma,\xi,\eta)=
\mbox{res}_{\gamma=\infty} \omega^k \partial_\xi \log \psi_1(\gamma,\xi,\eta),
\end{equation}

\begin{equation}
\label{SG-3.2.4}
\frac{\partial A^k(\vec\gamma)}{\partial \eta}=
-\sum_k \mbox{res}_{\gamma=\gamma_k(\xi,\eta)} \omega^k
\partial_\eta \log \psi_1(\gamma,\xi,\eta)=
\mbox{res}_{\gamma=0} \omega^k \partial_\eta \log \psi_1(\gamma,\xi,\eta).
\end{equation}

Calculating residues in (\ref{SG-3.2.3}), (\ref{SG-3.2.4}) we obtain:

\begin{equation}
\label{SG-3.3} \frac{\partial A^k(\vec\gamma)}{\partial \xi}=
2D^k_{g-1},
\end{equation}

\begin{equation}
\label{SG-3.4} \frac{\partial A^k(\vec\gamma)}{\partial \eta}= 2
\frac{D^k_{0}}{\sqrt{\prod\limits_{j=1}^{2g}E_j}},
\end{equation}

and

\begin{equation}
\label{SG-3.5} \frac{\partial A^k(\vec\gamma)}{\partial x}=
\frac12\left(D^k_{g-1}+
\frac{D^k_{0}}{\sqrt{\prod\limits_{j=1}^{2g}E_j}}\right)=
-\frac{1}{2\pi}\oint\limits_{b_k}dp,
\end{equation}
where $dp$ denotes a meromorphic differential with zero
$a$-periods and two second order poles at $0$, $\infty$ such, that

\begin{equation}
\label{SG-3.5.1}
dp=\left\{
\begin{array}{l}
\left(\frac{\displaystyle 1}{\displaystyle 8\sqrt{\lambda}}+
o\left(\frac{\displaystyle 1}{\displaystyle\lambda}\right)\right)d\lambda
\ \ \mbox{as} \ \ \lambda\rightarrow\infty,
\\ \\
\left(\frac{\displaystyle 1}{\displaystyle 8\lambda \sqrt{\lambda}}+
o\left(\frac{\displaystyle 1}{\displaystyle \lambda}\right)\right)d\lambda
\ \ \mbox{as} \ \ \lambda\rightarrow
 0
\end{array}
\right.
\end{equation}

Therefore the density of topological charge is equal to the
following expression

\begin{equation}
\label{SG-3.6}
\bar n=\frac12 \ \sum\limits_{k=0}^{g-1}
\left(D^k_{g-1}+
\frac{D^k_{0}}{\sqrt{\prod\limits_{j=1}^{2g}E_j}}\right)n_k,
\end{equation}
where $n_k$ denote the topological charge of the image of the
$k$-th basic cycle for the given topological component. By definition
$n_k$ is an integer.

To calculate $n_k$ for a given topological type consider a basic
cycle ${\cal A}_k$ on the real component of Jacoby torus.
represented by the closed curve $B_k$ on this torus. The image of
this cycle
 in the
surface $\Gamma$ (i.e. image of it under the inverse Abel map) is
a closed oriented curve $C_k$ (may be consisting from several
connected components). This curve is presented geometrically as a
union of the divisor points $D(t_k)=\sum\gamma_j(t_k)$ for all
values of the corresponding ''time'' $t_k$ along this basic cycle.
It is homological to the cycle $a_k\in H_1(\Gamma,Z)$. The
projection of the last curve $C_k$ on the $\lambda$-plane we
denote $\tilde{C}_k$. This curve does not touch the closed
segments on the real line $[-\infty,E_{2m}],\ldots,[E_3,E_2],
[E_1,0]$.

\begin{lemma}
\label{SG-L4.1}
The topological charge $n_k$ along the cycle ${\cal A}_k$ is given by the
following formula:
\begin{equation}
\label{SG-3.6.1}
n_k=\tilde C_k \circ {\Bbb R}_-,
\end{equation}
where $\tilde C_k$ is the projection of $C_k$ to the
$\lambda$-plane, ${\Bbb R}_-$ is the negative part of the real
line with the standard orientation, and $\circ$ denotes the
intersection index of  curves on the $\lambda$-plane.
\end{lemma}

This Lemma follows directly from formula (\ref{SG-13}).

Assume now, that the cycles $b_1$, $b_2$, \ldots, $b_m$ are the ovals
lying over the segments $[E_2,E_1]$, $[E_4,E_3]$, $[E_{2m},E_{2m-1}]$
respectively. It is easy to check, that the function
$\mu$ at the sheet $G_+$ is negative at the intervals $[E_{4k-2},E_{4k-3}]$
and positive at the intervals $[E_{4k},E_{4k-1}]$.

\begin{center}
\mbox{\epsfxsize=10cm \epsffile{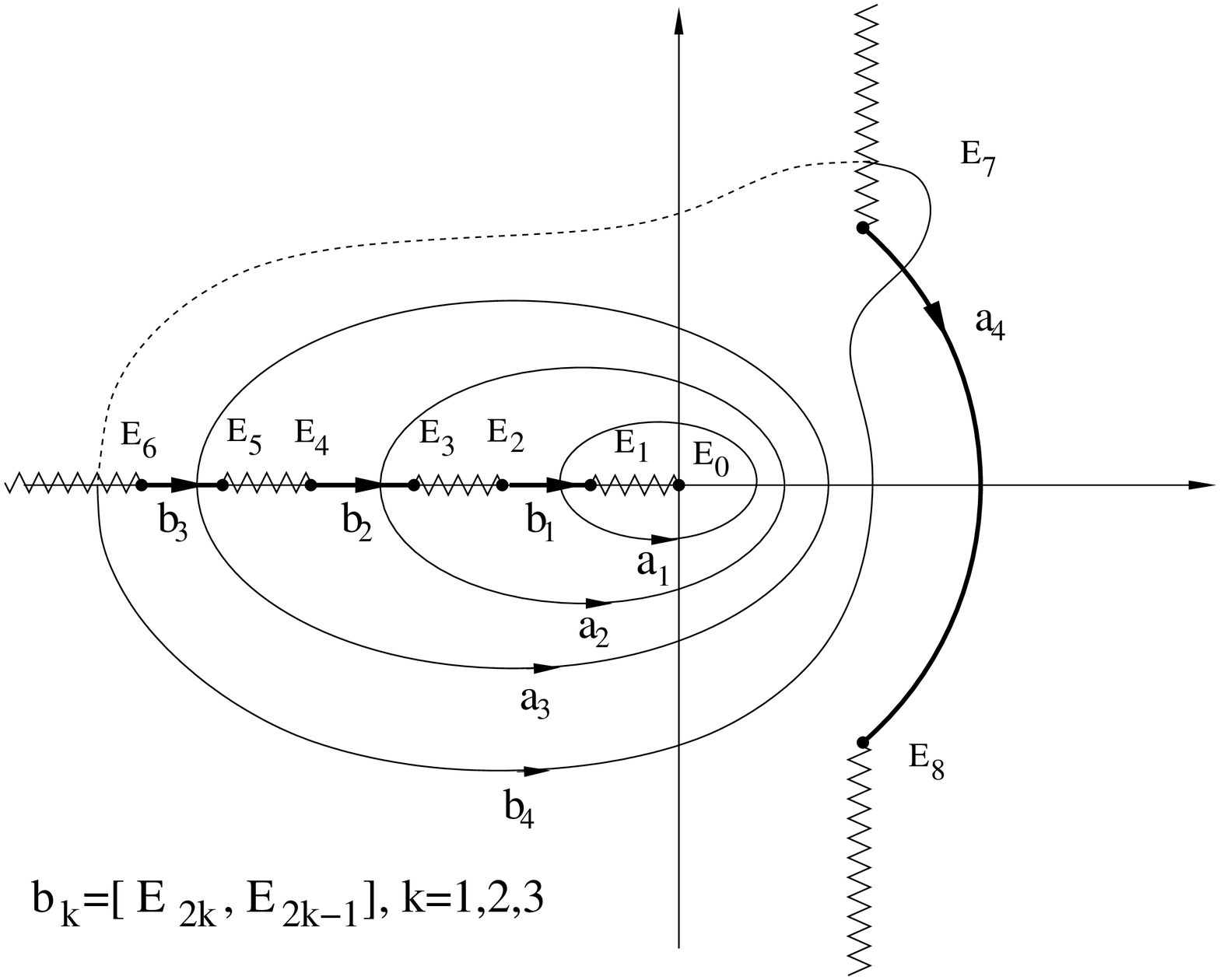}}

Fig 4.
\end{center}

It follows from Lemma~\ref{SG-L2}, that the projection-curve
$\tilde C_k$ can cross the half-line ${\Bbb R}_-$ only strictly
inside the intervals $[E_2,E_1]$, $[E_4,E_3]$,
$[E_{2m},E_{2m-1}]$. Therefore any time when the curve $\tilde
C_k$ crosses the half-line ${\Bbb R}_-$, the cycle $C_k$ on the
Riemann surface $\Gamma$ crosses one of the cycles $b_1$, $b_2$,
\ldots, $b_m$. The orientation of the cycles $b_l$, $l=1,\dots,m$
coincides with the orientation of ${\Bbb R}_-$ at the sheet $G_+$;
 they are opposite at the sheet $G_-$. It follows from
Lemma~\ref{SG-L2.1}, that the cycle $C_k$ always crosses $b_l$ in
the sheet $G_+$ if $(-1)^{l-1}s_l>0$; it crosses $b_l$ in the
sheet $G_-$ if $(-1)^{l-1}s_l<0$. Therefore we proved

\begin{lemma} The topological charge along the cycle ${\cal A}_k$
on the real torus in the Jacobian variety is equal to the
intersection index of the corresponding curve $C_k$ with the {\bf
Counting Cycle of the real component} on the Riemann surface
$\Gamma$
\begin{equation}
\label{SG-3.6.2} n_k=C_k\circ \left(\sum\limits_{l=1}^m (-1)^{l-1} s_l
b_l\right)
\end{equation}
\end{lemma}
Let us mention that the cycle $C_k$ is homotopic to $a_k$ on
$\Gamma$. Therefore $C_k\circ b_l=\delta_{kl}$ and
\begin{equation}
\label{SG-3.6.3}
\left\{
\begin{array}{ll}
n_k=(-1)^{k-1} s_k & \mbox{for} \ \ k\le m \\
n_k=0 & \mbox{for} \ \ k> m.
\end{array}
\right.
\end{equation}
Combining (\ref{SG-3.6}) with (\ref{SG-3.6.3}), we obtain our main formula
(\ref{SG-3.2}).

The Theorem is proved.

Let us present here  a second method for calculating the
topological charges along the basic cycles. By Lemma~\ref{SG-L1.2}
the topological components depend continuously on the branching
points. Assuming that the admissible polynomial is given by
(\ref{SG-2.6}) and all points of the divisor $D$ lie in the upper
half-plane we obtain a family of solutions continuously depending
on the branching points. The motion of an individual zero point
$\gamma_j$ with any $2\pi$-cyclic time $\tilde{t}_s $
corresponding to the basic cycle ${\cal A}_s$ on the real torus
does not depend continuously on the initial data, because the
divisor points may collide with each other; the Dubrovin equations
became singular in the collision point. The set of singular
divisors has real codimension 2. But we deduce from the
$\Theta$-functional formulas  that all symmetric functions of the
divisor points pass through this singularity continuously.
 All real
solutions are nonsingular. Therefore the topological charges along
the basic cycles are integers continuously depending on the
branching points. We conclude that  they are constant.

For the proof of (\ref{SG-3.6.3}) it is sufficient to calculate
the number $n_k$ for one spectral curve $\Gamma_k$. Let $k\le m$. Consider
the following configuration of the branching points for this curve $\Gamma_k$:

\begin{equation}
\label{SG-3.6.4}
\left\{
\begin{array}{lll}
E_{2k}=-k-\epsilon_1 & E_{2k-1}=-k+\epsilon_1 & \\
E_{2j}=-j-\epsilon_2 & E_{2j-1}=-j+\epsilon_2 & \mbox{as} \ \ j\le m, \ j\ne k \\
E_{2j}=j-i\epsilon_2 & E_{2j-1}=j+i\epsilon_2 & \mbox{as} \ \ j> m,
\end{array}
\right.
\end{equation}
where $\epsilon_2\ll\epsilon_1\ll 1$.

The cycle ${\cal A}_k$ can be realized by the following 1-parametric family
of polynomials $P_{g-1}(\lambda,\beta)$:
\begin{equation}
\label{SG-3.6.5}
P_{g-1}(\lambda,\beta)=(-1)^{k+g-m-1} s_k \beta \prod\limits_{j=1}^{g-1}
(\lambda-r_j),
\end{equation}
where
\begin{equation}
\label{SG-3.6.6}
\begin{array}{ll}
r_{j}=-j-(-1)^{k+j}s_js_k 2|j-k|\frac{\epsilon_2}{\epsilon_1}\sqrt{\frac{k}{j}}
 & \mbox{as} \ \ j<k \\
r_{j}=-j-1-(-1)^{k+j}s_js_k 2|j+1-k|\frac{\epsilon_2}{\epsilon_1}
\sqrt{\frac{k}{j+1}}
& \mbox{as} \ \ k\le j < m \\
r_{j}=j+1 & \mbox{as} \ \  m\le j <g,
\end{array}
\end{equation}
and the parameter $\beta$ runs through the interval
$[\frac{\epsilon_1}{\sqrt{k}}(1+o(1)),2\sqrt{k}(1+o(1)]$.

Elementary estimates show, that system (\ref{SG-2.2}) has a pair
of roots located in a neighborhood of  the circle $|\lambda|=k$
and all other roots form complex conjugate pairs located in small
neighborhoods of the point $r_n$. The distinguished pair of roots
lie in the sheet $G_+$ if $(-1)^{k-1}s_k>0$ and in $G_-$ if
$(-1)^{k-1}s_k<0$. Therefore we proved (\ref{SG-3.6.3}) for $k\le
m$.

To calculate $n_l$ with $l>m$ it is sufficient to consider the same
spectral curve as above. Let us assume $\beta=1$ and consider a small variation
of the point $r_{l-1}$. It gives us the cycle $a_l$. Variation of
the function $u(\xi,\eta)$ along this cycle is very small, therefore $n_l=0$.

The second proof of the Theorem is completed.

\section{Appendix 1. The Cherednik differential for the generic periodic real
SG solutions.}

Let us explain here why the sufficient reality conditions from the
Section~\ref{SG-Sect2.3} are in fact necessary for the generic
smooth $x$-periodic  solutions. Let us remind, that in the SG
theory we call a potential $u(x,t)$ periodic if the function
$\exp\{u(x,t)\}$ is periodic. Therefore the operator
$$
L(\lambda)=\partial_x-1/4(U+V)
$$
has a Bloch solution
$$
L(\lambda)\Psi=0,\ \ \hat{T}\Psi_{\pm}(\lambda,x,t)=
\Psi_{\pm}(\lambda,x+T,t)=\exp\{\pm ip(\lambda)T\}
\Psi_{\pm}(\lambda,x,t)
$$
where the differential $dp$ on the Riemann surface has been
defined above in the finite-gap case. This formula works for all
complex values of $\lambda$ except branching points where the
shift operator $\hat{T}$ has the Jordan cells. The Bloch
vector-function $\Psi_{\pm}(\lambda,x)$ is in fact a meromorphic
one-valued vector-function $\Psi(\gamma,x)$ on the Riemann surface
$\Gamma$ whose points are exactly the pairs
$\gamma=(\lambda,\pm)\in\Gamma$ or $\gamma=(\lambda,\mu)$ for
$\mu=\exp\{ipT\}$. This surface is obviously 2-sheeted, i.e.
hyperelliptic. For the real solutions this surface has a pair of
commuting involutions (holomorphic and antiholomorphic):
$$
\sigma(\lambda,+)=(\lambda,-), \ \ \mbox{or} \ \
\sigma(\lambda,\mu)=(\lambda,\mu^{-1}),
$$
$$
\tau(\lambda,\mu)=(\bar\lambda,\overline{\mu^{-1}}).
$$
The action of $\sigma\tau$ on the Bloch solutions is given by the
following explicit formula
\begin{equation}
\label{SG-4.1} \left(\begin{array}{c}
\psi_1(\sigma\tau\gamma,x,t) \\ \psi_2(\sigma\tau\gamma,x,t)
\end{array}\right)=\frac{1}{\overline{\psi_2(\gamma,0,0)}}
\left(\begin{array}{c}
\overline{\psi_2(\gamma,x,t)} \\
-\overline{\lambda\psi_1(\gamma,x,t)}.
\end{array}\right)
\end{equation}

{\bf Statement.} The Riemann surface $\Gamma$ always has the
points $0,\infty$ as its branching points. For the real smooth
$x$-periodic SG solutions $u(x,t)$ this surface has no real
positive branching points. Let this surface be nonsingular. Then
we define a  ''Cherednik differential'' $\Omega$: by definition,
its  zeroes should coincide with the full set of poles of the
Bloch function with the proper normalization
\begin{equation}
\label{SG-4.2} \psi_{1,\pm}(\lambda,x,t)=1 \ \ \mbox{for} \ \
x=0,t=0,
\end{equation}
and their $\tau$-conjugate points. It has also simple poles in the
points $0,\infty$. This differential can be written in the
following form:
\begin{equation}
\label{SG-4.3}
\Omega=\frac{\psi_1(\gamma,0,0)\psi_2(\sigma\gamma,0,0)}
{\psi_1(\gamma,0,0)\psi_2(\sigma\gamma,0,0)-
\psi_1(\sigma\gamma,0,0)\psi_2(\gamma,0,0)}d\lambda.
\end{equation}

Let us check that the differential $\Omega$ defined by (\ref{SG-4.3})
is holomorphic outside the points $0$, $\infty$ and has the correct
set of zeroes.
We see, that the form $\Omega$ does not depend on the
normalization of the Bloch function $\Psi$. Therefore we may
assume, that any locally holomorphic normalization is used. Then the
denominator of formula (\ref{SG-4.3}) has zeroes exactly at the
points where the solutions $\Psi(\gamma,x,t)$ and
$\Psi(\sigma\gamma,x,t)$ are linearly dependent, i.e. at the
branching points. The surface $\Gamma$ is nonsingular, therefore
these zeroes are simple and the form $\Omega$ has no singularities
at the branching points.

The zeroes of the form $\Omega$ are the zeroes of the first
component of the Bloch function $\psi_1(\gamma,0,0)$ plus the
zeroes of the second multiplier $\psi_2(\sigma\gamma,0,0)$. It follows from
formula (\ref{SG-4.1}) that zeroes of the function
$\psi_2(\sigma\gamma,0,0)$ are $\tau$-conjugate to the zeroes of
the function $\psi_1(\gamma,0,0)$. Taking into account that the poles of the
Bloch function normalized by the condition (\ref{SG-4.2}) are exactly the zeroes of
$\psi_1(\gamma,0,0)$, we complete the proof.

Let $\lambda$ be a real positive number. Then the restriction of
the shift operator $\hat T$ to the zero eigenspace of the
operator $L(\lambda)$ is unitary. Therefore  $\hat T$ and has no Jordan
cells, and $\lambda$ is not a branching point.

\section{Appendix 2. Numerical simulations.}

 During the work on this problem the following
numerical simulations were made:
\begin{enumerate}
\item The topological charges along the basic cycles on the real part of the
Jacoby torus were calculated for all topological types.
\item The densities of topological charge for the $x$-dynamics and for the
$\xi$-dynamics were directly calculated.
\end{enumerate}

For numerical calculation of topological charges $n_k$ along the
basic cycles we used the following approach. Assume, that the Baker-Akhiezer
function $\Psi(\gamma,\xi,\eta)$ has $2g$ additional essential singularities
at the points $(\lambda,\mu)=(\beta_s,\pm\sqrt{R(\beta_s)})$ with the following
asymptotics:
\begin{equation}
\label{SG-5.1}
\Psi(\gamma,\xi,\eta)=O(1)e^{\displaystyle -i\frac{\displaystyle \pm\sqrt{R(\beta_s)}}
{\displaystyle \lambda-\beta_s} \tilde t_s}
\end{equation}
where $\beta_s$, $s=1,2,\ldots,g$ are some real numbers. It is rather easy to check,
that these flows commute with the SG-dynamics. They correspond to the motion along
straight lines in the real components of the Jacoby torus:
\begin{equation}
\label{SG-5.2} \frac{\partial A^k(\vec\gamma)}{\partial\tilde
t_{s}}=W_s^k= 2\sum\limits_{l=0}^{g-1} D^k_{l}\beta^l_s.
\end{equation}
These flows are non-local in terms of the Sine-Gordon potential $u$, but the corresponding
Dubrovin equations are well-defined and have rational right-hand sides:
\begin{equation}
\left\{
\begin{array}{l}
\frac{\displaystyle \partial\lambda_l}{\displaystyle \partial\tilde t_{s}}
=-2i\ \frac{\displaystyle \mu_l\prod\limits_{j\ne l}(\beta_s-\lambda_j)}
{\displaystyle \prod\limits_{j\ne l}(\lambda_l-\lambda_j)}
\\ \\
\frac{\displaystyle \partial\mu_l}{\displaystyle \partial\tilde t_{s}}=-i\
\frac{\displaystyle R'(\lambda_l) \prod\limits_{j\ne l}(\displaystyle {\beta_s}-\lambda_j)}
{\displaystyle \prod\limits_{j\ne l}(\displaystyle \lambda_l-\lambda_j)}
\\ \\
\frac{\displaystyle \partial u}{\displaystyle \partial\tilde t_{s}}=-2\
\sum\limits_{l}\frac{\displaystyle \mu_l \prod\limits_{j\ne l}
(\displaystyle \beta_s-\lambda_j)}
{\displaystyle \lambda_l\prod\limits_{j\ne l}(\displaystyle \lambda_l-\lambda_j)}.
\end{array}\right.
\label{SG-5.3}
\end{equation}

To avoid singularities in the flows (\ref{SG-5.3}) generated by collisions of
divisor projections $\lambda_k(\xi,\eta,\tilde t_s)$ with the points $\beta_s$ we have
chosen the points $\beta_s$ in the forbidden intervals:
\begin{equation}
\label{SG-5.4}
\beta_s=\frac12 (E_{2s-1}+E_{2s-2}), \ \ \
s=1,\ldots,g.
\end{equation}

It follows from equation (\ref{SG-5.2}), that the linear combination of
non-local flows
\begin{equation}
\label{SG-5.5}
\frac{\partial}{\partial\nu_k} =\sum\limits_{l=1}^g (W^{-1})_k^l
\frac{\partial}{\partial \tilde t_l}
\end{equation}
where $W^{-1}$ denote the inversion of matrix $W$,
corresponds to the motion along the basic cycle ${\cal A}_k$ in the real
torus. Integrating (\ref{SG-5.5}) numerically we obtained the
topological charges along the basic cycles.

\section{Appendix 3. Fourier Transform  on the Riemann Surfaces:
The continuous
analog of the Krichever-Novikov bases.}

Let us invent here a continuous analog of the so-called
``Krichever-Novikov bases'' on the Riemann surfaces defined and
used by the authors for the needs of the quantum string theory  in
the late 80s--see the first work in this series \cite{KN}. These
bases have been considered as a natural analog of the
Fourier-Laurent bases in complex analysis.

Let us consider  the following set of data: a nonsingular Riemann
surface $\Gamma$ of the genus $g$ with marked point
$\infty\in\Gamma$ and selected local parameter near this point
$z=k^{-1}, z(\infty)=0$. We construct a function
$\psi^0=\psi(P,x)$ holomorphic on $\Gamma\backslash \infty$  and
exponential near the infinite point:
$$\psi(z,x)=k^{g}\exp\{kx\}(1+\sum_{i>0} \eta_i(x)k^{-i}), \ \ \eta_0=1$$
It is obviously a partial case of the general Krichever
KP-construction for the case where the divisor of poles is equal
to $g\infty=D$. This case leads  to the singular solutions of the
 KP-type systems. Therefore it is out of use in the physical soliton
theory. However, the examples of this kind appeared in the works
of XIX century as the so-called Lam\'e operator. It  appears as a
result of  the separation of variables in the Laplace-Beltrami
operator on the 3-axis ellipsoid. The spectral theory on the
closed segment between two singularities has been studied by the
classical people like Hermit and others:

{\bf Example.} For the genus $g=1$ we have:
$$\psi(z,x)=\frac{\sigma(z-x)}{\sigma(z)\sigma(x)}\exp\{x\zeta(z)\}$$
This function satisfies to the Lam\'e equation
$$(\partial_x^2-2\wp(x))\psi= \wp(z)\psi(z,x)$$

We construct also a differential form $\psi^{1}=\phi(P,x)dz$ on
$\Gamma$ such that it is holomorphic outside of the infinite point
and has following asymptotics near it:
$$\phi(P,x)dz=z^{g-2}\exp\{-kx\}(1+o(1))dz.$$

Consider   a $x$-independent second kind differential $dp$ on the
surface $\Gamma$ with the unique pole of the second order in the
point $\infty\in\Gamma$ and with asymptotics: $$dp=dk(1+o(1)),\ \
k=z^{-1}\rightarrow\infty$$ We require that this differential has
following property for all 1-dimensional cycles on the Riemann
surface: $$\Re\oint_Adp=0,\ \ A\in H_1(\Gamma,Z)$$ Therefore the
function $\Re p=\tau$ is an one-valued real harmonic function
$\tau(P)$ on the Riemann surface with the unique pole in the point
$\infty$. This function will be called ``time'' by the analogy
with the string theory \cite{KN}. All its levels enter the point
$\infty$ twice.

 {\bf Example.} For the genus equal to 1 $g=1$  and
real algebraic curves we have the periods  $2\omega\in R$ and
$2i\omega'\in iR$. The function defined above has a form:
 $$
p(z)=\zeta(z)-\frac{\eta}{\omega} z, \ \
dp(z)=\left(-\wp(z)-\frac{\eta}{\omega} \right)dz$$
 where
$\eta=\zeta(\omega)$ (see the book \cite{BE}).

{\bf\Large Problem.} How to construct Harmonic Analysis on the
levels $\tau=const$ and in the complex domains $c_1<\tau<c_2$
using the special Baker-Akhiezer function $\psi(P,x)$ described
above? Which multiplicative properties have the basic functions
$\psi(P,x)=\psi_x(P)$ depending on $x$ as parameter?

{\bf Proposition.}The Fourier-Laurent decomposition formula for
the function $f(P)$
 has a form:
$$f(P)=\int\left \{ 1/2\pi\oint_{\tau=const}\phi(P',x)f(P')dz
\right \} \psi(P,x)dx$$
 here the points $P,P'$ belong to the level
$\tau=const$ or to the domain $c_1<\tau<c_2$. All connectivity
components should enter this integral. This formula valid for the
class of functions with the proper decay near $\infty$. We shall
publish more precise statement later.

  We consider here the subscript $x$ as a continuous analog of
the discrete $n\in Z$ from the work \cite{KN}. Therefore we expect
to have here some natural analog of the ``almost grading
property'' discovered in \cite{KN}:

\begin{theorem}.
Let $x,y\neq 0$. There exists a differential operator $L$ in the
variable $x$ of the order $g$ with coefficients dependent on the
both variables $x,y$ such that the following {\bf Almost Graded
Commutative Associative Ring Structure} is defined by the formula
$$\psi(P,x)\psi(P,y)=L\psi(P,x+y)$$
$$L=\partial_x^g+[\eta(x)+\eta(y)-\eta(x+y)]\partial_x^{g-1}+\ldots
$$ The coefficients of this operator have poles  for the points
$x=0$ or $y=0$.
\end{theorem}

In the example above where $g=1$ we have $$
L=\partial_x-(\zeta(x)+\zeta(y)-\zeta(x+y)) $$

In the next work we shall present more detailed description of
this idea including the higher tensor generalization $\psi^k$, the
decomposition theorems for the functions and tensors on the
contours $\tau=const$ and domains $c_1<\tau<c_2$ following the
program of the work \cite{KN}.

\end{document}